\documentclass[twocolumn,trackchanges,twocolappendix]{aastex631}

\usepackage{soul}
\usepackage{xcolor}
\definecolor{dkblue}{RGB}{54, 86, 169}
\newcommand{\EQ}[1] {equation~(\ref{#1})}
\newcommand{\SEC}[1] {Section~\ref{#1} }

\newcommand{\FIG}[1] {Figure~\ref{#1} }

\newcommand{\C}{\textsf{\textbf{C}}}

\newcommand{\PA}{PSR~B0355$+$54}
\newcommand{\PB}{PSR~B0329$+$54}

\usepackage{amsmath}
\begin{document}

\title{A  Linear Decomposition Method to Analyze and Study Pulsar Mode Changes}

\author{Longfei Hao}
\affiliation{Yunnan Observatories, Chinese Academy of Sciences, 650216 Kunming, China}
\affiliation{Key Laboratory for the Structure and Evolution of Celestial Objects,\\
Chinese Academy of Sciences, 650216 Kunming, China}

\author[0000-0001-8021-4559]{Zhixuan Li}
\affiliation{Yunnan Observatories, Chinese Academy of Sciences, 650216 Kunming, China}
\affiliation{Key Laboratory for the Structure and Evolution of Celestial Objects,\\
Chinese Academy of Sciences, 650216 Kunming, China}

\author{Faxin Shen}
\affiliation{Yunnan Observatories, Chinese Academy of Sciences, 650216 Kunming, China}
\affiliation{Key Laboratory for the Structure and Evolution of Celestial Objects,\\
Chinese Academy of Sciences, 650216 Kunming, China}

\author[0000-0001-5662-6254]{Yonghua Xu}
\affiliation{Yunnan Observatories, Chinese Academy of Sciences, 650216 Kunming, China}
\affiliation{Key Laboratory for the Structure and Evolution of Celestial Objects,\\
Chinese Academy of Sciences, 650216 Kunming, China}

\author[0009-0000-8142-8612]{Yuxiang Huang}
\affiliation{Yunnan Observatories, Chinese Academy of Sciences, 650216 Kunming, China}
\affiliation{Key Laboratory for the Structure and Evolution of Celestial Objects,\\
Chinese Academy of Sciences, 650216 Kunming, China}

\author{Kejia Lee}
\affiliation{Department of Astronomy,School of physics, Peking University, Beijing, 100871}
\affiliation{National Astronomical Observatory, Chinese Academy of Sciences, Beijing 100101, China}
\affiliation{Beijing Laser Acceleration Innovation Center, Huairou, Beijing, 101400, China}

\author{Qingzheng Yu}
\affiliation{Department of Astronomy, Xiamen University, Xiamen, 361005, China}

\author{Hongguang Wang}
\affiliation{School of Physics and Electronic Engineering, Guangzhou University, Guangzhou, 510006, China}
\correspondingauthor{Longfei Hao, Zhixuan Li}
\email{haolongfei@ynao.ac.cn, lzx@ynao.ac.cn}


\begin{abstract}
In this paper, we present the linear decomposition method (LDM), which we developed to detect and analyze pulsar profile variations and mode changing behaviour. 
We developed LDM utilizing the likelihood function approach assuming the Gaussian noise. The LDM projects pulse profiles onto significance-ordered orthonormal vector bases. We show that the method is similar to the principal component analysis (PCA), but LDM can handle more general situations.
We use simulated dataset and data from the Kunming 40-m radio telescope to demonstrate the application of the LDM. We found that the LDM successfully identified mode changes for well-known mode-changing \PB~ and found a continuous pulse profile evolution for \PA~. We also show that the LDM can be used to improve the timing precision for mode changing \PB.

\end{abstract}


\section{Introduction} \label{sec:intro}

Radio pulsars are a class of rapid spinning neutron stars radiating in the radio band. Due to the star's rotation, the collimated radiation beam manifests itself as pulsed emission for distant observers. It has been widely observed that most radio pulsars have stable average pulse profiles 
that are formed by integrating usually at least a few hundred to a few thousand individual pulses (depending on a pulsar).
However, some pulsars exhibit mode-changing (or mode-switching) phenomena, whereby they switch between two or more shapes of average profiles. 
This phenomenon has been detected in a few dozen pulsars \citep{Lyne1971,MSF80,MGSB81,FWM81,LHK10} since its first discovery in PSR B1237+25 by \citet{Backer1970}. 
It has also been found that mode changes also occurs in X-rays \citep{HHK13,MKT16}, in addition to radio emissions. 
Discoveries of correlation between pulsar spin-down rate and mode-changing activities \citep{LHK10} further show that mode change is a possible result of the magnetosphere change \citep{Timokhin10}, which creates the link between the pulse profile morphology and pulsar dynamics. 

For most of the reported mode-changing pulsars, the mode-changing events, i.e. the abrupt shape changes of pulses, can be detected by the naked eye.
For studies on typical mode-changing pulsars and not involving a large amount of data, it is possible to identify the mode-changing events simply by `visual inspection'.
However, to systematically identify, analyze, and quantify the mode-changing phenomena with a large number of pulses, the `visual inspection' would be an exhausting process and the statistics can be biased by human factors. 

If the profile template is not available, one of the methods to find the pulse profiles with different radiation modes is the principal component analysis (PCA) \citep{B1991PhDT}.  PCA characterizes the pulse profile shape variations by forming a set of
orthonormal basis vectors along which the profile variance is maximized.
PCA can also be viewed as a low rank approximation of pulse profiles. For example, if the pulse profile is constant, then only one profile (one basis vector) is needed to describe the profile,
i.e. the pulse profile matrix\footnote{Pulse profile matrix is defined that each row of the matrix is one observed pulse profile with the column corresponding to the pulse phase. } becomes rank-1.  

In this paper, complementary to the PCA method, we developed a linear decomposition method (LDM) to detect and analyze the shape variability in a pulsar's pulse profile. Comparing with the PCA method, the LDM is also a weighted low-rank approximation (WLRA). It more general than the PCA method and extra constraints can be implemented. Further more, when only a limited number of modes are under study, the computational cost of the current method is much smaller than the PCA with full singular value decomposition (SVD).

In Section 2, we describe the LDM algorithm and its implementation and testing with simulated data. Our demonstrative application of LDM to Kunming 40 meter radio telescope data of the two mode-changing sources \PB~ and \PA~ are presented in Section 3.  Discussions and conclusions are presented in Section 4.

\section{linear decomposition method of pulse profiles}
\label{sec:method}

As the integration time of a given pulse profile may be longer than the duration of certain pulse profile mode, all the integrated pulses are regarded as the mixture of characteristic pulse profiles. If no mode change, the `mixture' is of only one mode. If the integration is long enough, the integrated pulse profile contains all the possible modes. Clearly, the weight of each mode in the mixture is the total energy pulsar radiated in the given mode. 

Thus, any pulse profile is a linear combination of a set of basis 
profiles corresponding to different emission modes, i.e.
\begin{eqnarray}
		p_{ji} = \sum_{k=1}^{M} \alpha_{jk}f_{ki} + \varepsilon_{ji} 
		\,.\label{eq:model}
\end{eqnarray}
where $p_{ji}$ represents the measured pulse profile at the $i$-th phase bin for 
the $j$-th sub-integration. $f_{ki}$ is the basis profile of the $k$-th mode.  
$k$ runs from 1 to $M$, the total number of modes.  $\alpha_{jk}$ represents the 
weight of the $k$-th mode profile base for the $j$-th sub-integration. 
$\varepsilon_{ji}$ is the decomposition residuals for the $j$-th sub-integration.  
For convenience, we employ the vector-matrix notation that
\begin{eqnarray}
{\rm \bf P} &\equiv& \left[p_{ji}\right]_{\rm n_{sub}\times n_{bin}}\,, \\
{\rm \bf A} &\equiv& \left[\alpha_{jk}\right]_{\rm n_{sub}\times M}\,, \\
{\rm \bf F} &\equiv& \left[f_{ki}\right]_{M\times\rm n_{bin}}\,, \\
{\boldsymbol \varepsilon} &\equiv& \left[n_{ji}\right]_{\rm n_{sub}\times n_{bin}}\,,\\
{\it \bf f}_k &\equiv& \left(f_{k1}, f_{k2},..f_{ki},.., f_{k\, \rm 
n_{bin}}\right)^T\,.
\end{eqnarray}
Here, $n_{\text{sub}}$ is the total number of sub-integration and
$n_{\text{bin}}$ is the number of phase bins for each pulse profile.
With the matrix notation, \EQ{eq:model} becomes ${\rm
\bf P} = {\rm \bf AF} + {\boldsymbol \varepsilon}$.

For Gaussian noise, the likelihood function of the data is
\begin{eqnarray}
\Lambda \propto e^{ -\frac{1}{2} \sum_i \sum_j 
\left(\frac{p_{ji}-\sum_{k=1}^{M}\alpha_{jk} 
f_{ki}}{\sigma_{ji}}\right)^2}\,.
\end{eqnarray}
In the above likelihood function, the baseline offsets of pulse profile is not 
modeled, because we by-default subtracted the baseline according to the 
off-pulse data.
The maximum likelihood estimator for the weights $\alpha_{jk}$ and profile bases 
$f_{ki}$ can be obtained by solving following equations \begin{eqnarray}
 \frac{\partial \Lambda}{\partial \alpha_{jk}}&=&0 \,,\\
 \frac{\partial \Lambda}{\partial f_{ki}}&=&0\,,
\end{eqnarray}
which correspond to:
\begin{eqnarray}
\sum_{i=1}^{n_{\rm bin}} \left(p_{ji}-\sum_{m=1}^{M}\alpha_{jm} f_{mi}\right) 
\frac{f_{ki}}{\sigma_{ji}^2} & =& 0\,, \label{eq:parta}\\
\sum_{j=1}^{n_{\rm sub}} \left(p_{ji}-\sum_{m=1}^{M}\alpha_{jm} f_{mi}\right) 
\frac{\alpha_{kj}}{\sigma_{ji}^2} & =& 0\label{eq:partb}\,.
\end{eqnarray}
There are two properties are worthy-note. 1) the equations are nonlinear. Terms 
with multiplication of unknown variables, e.g. $\alpha_{jm} f_{mi}$ appeared.  
2) the maximal number of independent equations is the same as the number of 
unknown variables.  Here, \EQ{eq:parta} contains $n_{\rm sub}\times M$ 
equations, and \EQ{eq:partb} contains $n_{\rm bin}\times M$ equations. There are 
maximally $n_{\rm sub}\times M+n_{\rm bin}\times M$ \emph{independent} 
equations. The unknowns are $\alpha_{jk}$ and $f_{ki}$, i.e.  totally $n_{\rm 
sub}\times M+n_{\rm bin}\times M$ unknown values.  So, solutions to the above 
system can be found in principle, although the direct solution is rather 
challenging due to their nonlinearity. 

We employed following iterative method to obtain the solution. We began with a 
set of random initial for $\alpha_{jk}$ and solve the bases $f_{ki}$ and then
$\alpha_{jk}$ iteratively based on the following two steps: \begin{eqnarray}
\left(\begin{array}{c}
				 f_{1i}' \\
				 f_{2i}' \\
				 ... \\
				 f_{Mi}'
    \end{array}\right)=
    {\rm \bf R}^{-1}
    \left(\begin{array}{c}
		\sum_j \frac{p_{ji} \alpha_{j1}}{\sigma_{ji}^2} \\
		\sum_j \frac{p_{ji} \alpha_{j2}}{\sigma_{ji}^2} \\
		... \\
		\sum_j \frac{p_{ji} \alpha_{jM}}{\sigma_{ji}^2}
    \end{array}\right)\,,\label{eq:M1}
\end{eqnarray}
and
\begin{eqnarray}
    \left(\begin{array}{c}
				 \alpha_{j1}' \\
				 \alpha_{j2}' \\
         ... \\
				 \alpha_{jM}'
    \end{array}\right)=
    {\rm \bf Q}^{-1}
    \left(\begin{array}{c}
		\sum_i \frac{p_{ji} f_{1i}}{\sigma_{ji}^2} \\
		\sum_i \frac{p_{ji} f_{2i}}{\sigma_{ji}^2} \\
		... \\
		\sum_i \frac{p_{ji} f_{Mi}}{\sigma_{ji}^2}
    \end{array}\right)\,.\label{eq:M2}
\end{eqnarray}
The above scheme comes from solving \EQ{eq:parta} and (\ref{eq:partb}) 
iteratively, while treating another set of variable as known. Here, $\rm \bf R$ 
and $\rm \bf Q$ both are $M\times M$ matrices and their elements can be 
calculated by
\begin{eqnarray}
		R_{lk} &=&\sum_j
		\frac{\alpha_{jl}\alpha_{ik}}{\sigma_{ji}^2}\,,\\
		Q_{lk} &=&\sum_i
		\frac{f_{lj} f_{ki}}{\sigma_{ji}^2}\,. \label{eq:MQ}
\end{eqnarray}

However, if we view the above equations from a vector space perspective,
it becomes evident that the solution lacks uniqueness.
Here, we can consider all the pulse profiles as the vectors in an 
$n_{\rm bin}$-dimensional vector space. The profile bases span an 
$M$-dimensional vector space. Thus we are projecting the pulse 
profiles from a $n_{\rm bin}$-dimensional vector space 
into a vector space with smaller dimension, 
i.e. the $M$-dimensional vector space spanned by the profile bases. Equation 
(\ref{eq:parta}) and \EQ{eq:partb} is used to find the optimal $M$-dimensional
space (spanned by the vectors ${\vec f}_k$) to perform such projection and find the projection 
coefficients ($\alpha_{jk}$), which is equivalent to compute the weighted low-rank
approximation (WLRA) of the data matrix $\rm \bf P$ (e.g. \citealt{srebro2003wlra}).
Because any invertible linear transformation of ${\vec f}_{k}$ will span
the same $M$-dimensional vector space, which will not affecting the projection operation. 
As a result, \EQ{eq:parta} and (\ref{eq:partb}) does not directly leads to a unique 
solution. 

In practice, we start with one mode only, and use the iterative method to find 
the solution of such 1-rank projection. We then normalize the ${\vec f}_1$ such 
that ${\vec f}_1 \cdot {\vec f}_1=1$. Then we subtract the components of ${\vec 
f}_1$ from all measured pulse profiles, and repeat the steps. In other words, we 
find one profile base each time, subtract its contribution, and then find the 
next profile base. 
In this way, solution to each step is unique, and the orthonormal condition
of ${\vec f}_{k}$ is guaranteed automatically. Since the input pulse profiles in 
each step are all orthogonal to the profile bases obtained in previous steps.

At this stage, one may think that the LDM is equivalent to weighted PCA \citep{Brunton_Kutz_2022}, which also project the data onto an orthonormal basis  to minimize the residual variance. Indeed, the solution of our LDM will be the same as PCA, if all $\sigma_{ji}$ are identical. One of the major differences is that LDM allows for heteroscedasticity, i.e. the error $\sigma_{ji}$ can be different for each data point. For the PCA, usually the SVD algorithm is used, and error of all data point should be the same. 

We can also implement constraints not possible for the PCA. For example, we can make all the bases vector in the LDM be nonnegative-valued, by adding constraints of $f_{ki}\ge0$ during the iterative stage. For the PCA algorithm, this is not possible. Similarly, using likelihood-driver approach also allows for incorporating the prior (extra information about pulse profiles from other sources), if we interpret the the parameter inference in the Bayesian framework.

Another advantage of LDM is that we can stop at any step when we derive enough profile bases. One can monitor the convergence in the iteration of adding new basis and stop when the residual is bellow the error. In contrast, general PCA calculates the full-rank solution. Due to the limited number of modes of pulsar profiles, only a few bases will usually be sufficient for the study. Therefore, for analysis with a large number of pulses, LDM is much more computation- and memory-efficient than PCA.

Now, we turn to identifying the figure of merit to describe pulse shape. 
Any two profiles ${\vec p}_j$ and ${\vec p}_{j^\prime}$ will have exactly the same shape if and only if  
\begin{equation}
    \frac{\alpha_{j1}}{\alpha_{j^\prime 1}}
    = \frac{\alpha_{j2}}{\alpha_{j^\prime 2}}
    =...
    =\frac{\alpha_{jM}}{\alpha_{j^\prime M}}\,.
    \label{eq:same}
\end{equation}  

So, we define shape parameters, where the $k$-th shape parameter for profile ${\vec p}_j$ is
\begin{eqnarray}
		S_{j\,k} \equiv \frac{\alpha_{j~k+1}}{\alpha_{j1}}\,, (k = 1, 2,...,M-1).  
		\label{eq:Sk}
\end{eqnarray}
These shape parameters can uniquely determine a pulse shape, and can be used to
reconstruct the pulse profile ${\vec p}_j$ in the projected vector space by
\begin{eqnarray}
	{\vec p}_{j} = {\vec f}_{1}+\sum_k S_{j\,k}{\vec f}_{k+1}\,.
   \label{eq:reconstruct}
\end{eqnarray}
Clearly, the above reconstruction normalizes the pulse flux using the weights of 
the most common elements, i.e. the first mode. 

\subsection{Demonstration with simulated dataset}
\begin{figure}[ht!]
\centering
\includegraphics[width=\columnwidth]{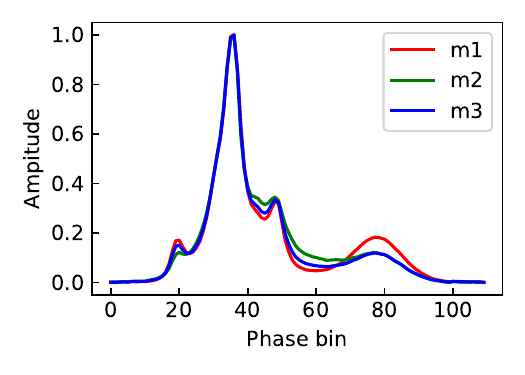}
\caption{Shapes of the modes used in generating the simulated profiles. They are obtained by mixing the integrated profiles of \PB~, \PA~ and a Gaussian profile as the tail component around the 80th phase bin. }
\label{fig:sim_modes}
\end{figure}

To demonstrate that the current LDM correctly detect shape variations including mode changes, we applied it to simulated dataset. The dataset consists of 2,000 profiles and includes three modes with modest differences. The three types of pulse profiles, \textbf{m1}, \textbf{m2}, and \textbf{m3}, are presented in \FIG{fig:sim_modes}. The number of profiles belonging to the three modes are 2, 598 and 1,400 respectively. 

\begin{figure}[ht!]
\centering
\includegraphics[width=\columnwidth]{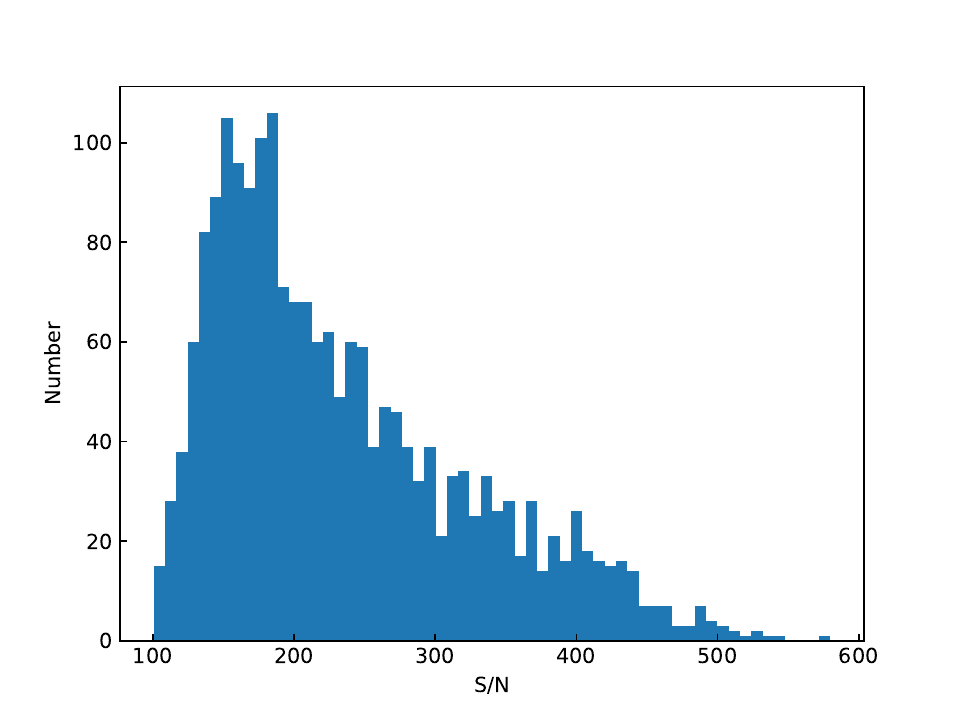}
\caption{Histogram of the simulated profile's S/N. }
\label{fig:sim_snr}
\end{figure}

\begin{figure}[ht!]
\centering
\includegraphics[width=\columnwidth]{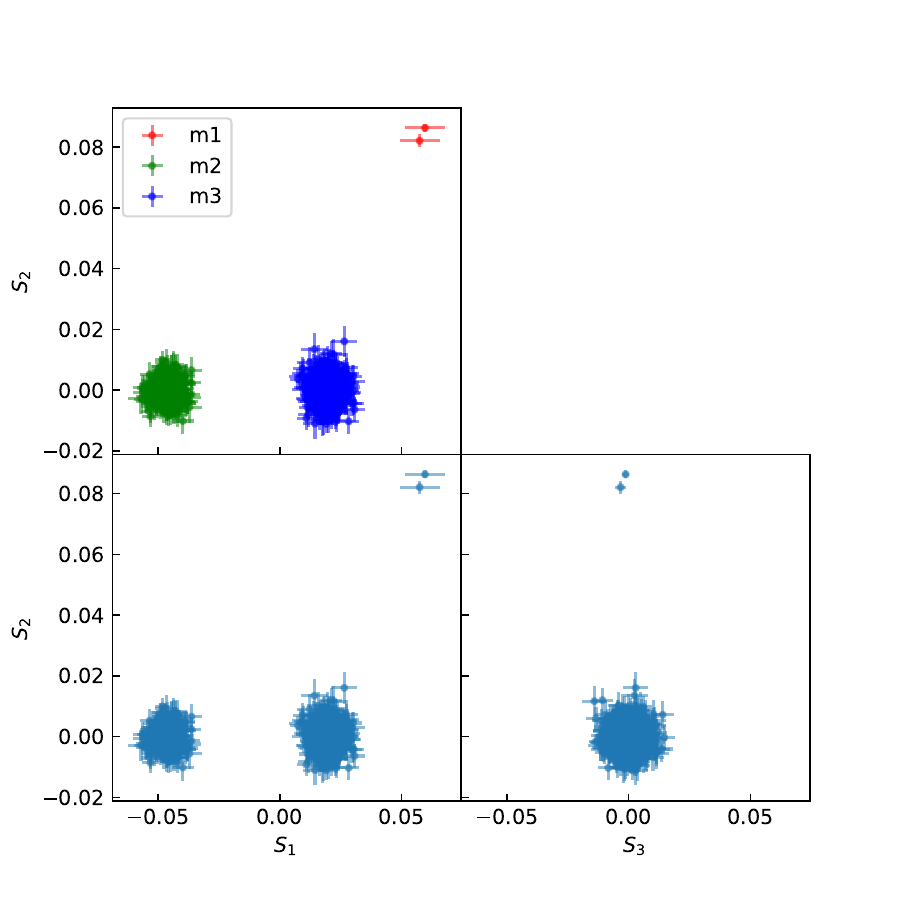}
\caption{The 2-D distributions of the shape parameters. The upper left panel is the $S_1$-$S_2$ distribution with the preset modes flagged with different colors. The two bottom panels are the $S_1$-$S_2$ and $S_3$-$S_2$ distributions, respectively. The errorbars for $S_1$...$S_3$ are estimated from the residuals.} 
\label{fig:sim_skdis}
\end{figure}

In the simulation, we added Gaussian noise to each profile, which made the profile S/N range from about 100 to 600. The histogram of S/N of our simulation is in \FIG{fig:sim_snr}. We then applied the LDM to the simulated profiles and derived the shape parameters described in \EQ{eq:Sk}. The distribution of the first three orders of $S_k$ is shown in \FIG{fig:sim_skdis}. The three modes are clearly identified and separated in the $S_1$-$S_2$ plot.

\section{Application the LDM to observation with Kunming 40-m telescope}
\label{sec:app2data}
We  apply the LDM to real data collected by the Kunming 40-meter radio telescope (KM40), which is operated by \emph{Yunnan Observatories} and located at $N25^{\circ}01'38'', E102^{\circ}47'45''$ \citep{hlf2010}.
Two pulsars are used here, i.e. \PB~ and \PA. \PB~ is well-known for its mode changing behaviour, while the mode changing events for \PA~was also reported before
(see \SEC{sec:0355} for more details). 

Our observations were carried out at S-band, 
centered at 2.256 GHz, with a bandwidth of 130 MHz. Signal digitizing and data recording were achieved by the 
\emph{Pulsar~Digital~Filter~Bank~4} (DFB4) system \citep{ferris2005256} developed by the 
\emph{Australia Telescope National Facility} (ATNF). Due to radio frequency interference, our right circular polarization data is corrupted. In the following analysis, we focus on the left circular polarization (LCP).


Data for \PB~ spans from January 2, 2016 to September 25, 2022 ($\sim 2,500$  days) and consists of over 2,600 observing sessions, with a total length of approximately 1,000 hours.
Data for \PA~ spans from 2014 May 16 to 2022 September 25 
($\sim3000$ days) and consists of over 1,200 observing sessions, with a total 
length of approximately 900 hours.

Most of the data was recorded with 30-second 
sub-integration time.
Each pulse contains 512 phase bins, and the summary of the 
observations can be found in Table \ref{tab:obssum}.  Pulsar parameters for 
folding were obtained from the ATNF Pulsar 
Catalog\footnote{http://www.atnf.csiro.au/research/pulsar/psrcat}
\citep{man2005}.

\begin{table}
	\centering 
	\caption{Summary of the observations on PSR
	\PB~ and \PA. The raw sub-integration time is 30 seconds, we
	further integrated them to 1/2/4/8 minutes for shape analysis. }
	\label{tab:obssum} \begin{tabular}{lr} %
		PSR B0329+54 & \\
		\hline
		\hline
        Number of sessions &  2622 \\
        Time span & MJD57389$-$59848\\
        Total length &  $\sim$1000 hours\\
        Sub-integration time  &  1/2/4/8 minute(s)  \\
        Frequency range  &  2190 - 2320 MHz  \\
        Frequency channel  &  1MHz$\times$130  \\
        Dedispersion reference frequency  &  2256 MHz  \\   
        Number of phase bins per period  &  512  \\
        \hline
        \hline
        \\
		PSR B0355+54 & \\
		\hline
		\hline
        Number of sessions &  1265 \\
        Time span & MJD56793$-$59847\\
        Total length &  $\sim$900 hours\\
        Sub-integration time  &  1/2/4/8 minute(s)  \\
        Frequency range  &  2190 - 2320 MHz  \\
        Frequency channel  &  1MHz$\times$130  \\
        Dedispersion reference frequency  &  2256 MHz  \\   
        Number of phase bins per period  &  512  \\
		\hline
        \hline
	\end{tabular}
\end{table}

Due to the presence of severe radio frequency interference (RFI) in most of the right 
circular polarization (RCP) data, we only used the left circular polarization 
(LCP) data in this study. We adopted a two-step approach to mitigate 
the RFIs. Firstly, we removed narrow channel RFIs by 
integrating the data across the span of given observing session and discarding 
the channels with spikes higher than 5-$\sigma$ of the bandpass. Secondly, we 
mitigated time-domain RFIs by removing any sub-integration where the 
signal-to-noise ratio (S/N) dropped below 5. After RFI removal, we further fold the 
30-second sub-integrations to create integrated profiles with integration times of 1, 
2, 4, and 8 minutes each, for subsequent analysis.

\subsection{Demonstration with PSR~B0329$+$54}
\label{sec:0329}
\PB~ is an ideal target for the demonstration of the mode-changing phenomenon. It is one of the brightest pulsars in the northern sky, and its mode changes had been intensively studied . \cite{Lyne1971} first observed the shape changes in the pulse profile of \PB~ at 408 MHz. The mode changes also occurred in observations
at 2.7 GHz and 14.8 GHz\citep{hesse1973, bartel1978}.
\cite{cjl2011} performed long-term monitoring of this source and obtained that the time duration ratio of the abnormal mode was approximately 15\%, which is consistent with earlier results\citep{bartel1982, XSG95}.

For practical pulsar data, there are two extra data processing steps before using the LDM.
The first is to subtract the baseline.
The second step is to align all pulses in phase to a template that can be simply obtained by a high-S/N average profile.
This has to be done, because any misalignment of the pulse profiles will introduce fake shape variations when performing the LDM, even when the profiles have the exact same shape.
This alignment is achieved by shifting the data profile with the phase shift determined by using the Fourier method \citep{taylor1992}.

To reduce the effects of noise, we selected profiles with $S/N\ge10$, resulting in a total data length of about 
1,000 hours. To reduce the computational time, we further selected profiles with 
$S/N\ge50$, when calculating the profile bases.  This resulted in a dataset with 
a total length of about 480 hours and approximately 29,000 profiles, each with 
an integration time of about 1 minute.  
Our final data selection step is to use only the on-pulse data. The pulse width of \PB~ is less than 10\%. Selecting the on-pulse data will further reduce the computational time and save significant computational and memory resources.

\begin{figure}[ht!]
\centering
\includegraphics[width=\columnwidth]{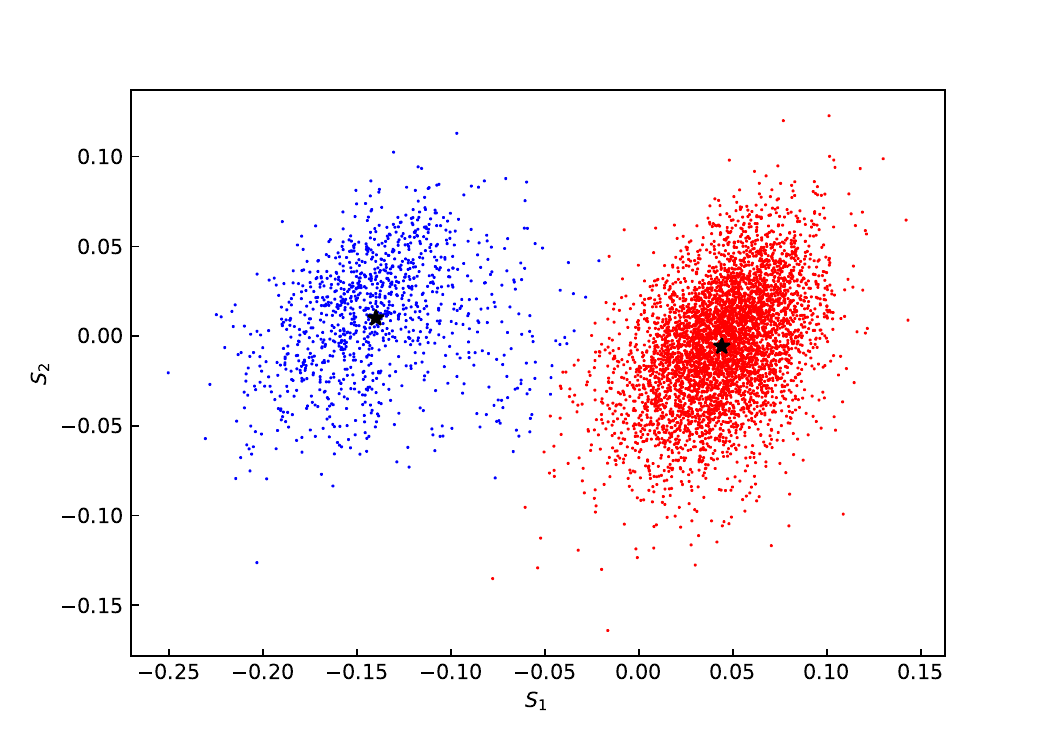}
\caption{Distribution of shape parameter of profiles of \PB.  
Here, we selected pulses with ${\rm S/N}>50$.  The GMM classifications are 
labeled using red and blue colors. The black stars are the centers of the 
two Gaussian components computed using the GMM method. To improve 
readability, we plotted 10\% of all data.
\label{fig:0329clss}}
\end{figure}

We apply the LDM to the selected data.  The two-dimensional distribution of the first two shape 
parameters is shown in Figure \ref{fig:0329clss}. We clearly detected two modes 
separated along the $S_1$ axis. We then used the Gaussian mixture model (GMM) 
\citep{press2007numerical, kjl2012} to separate the two populations. The centers 
of the Gaussian components are used to recover the pulse profiles corresponding to the two 
modes, i.e.
\begin{eqnarray}
	 \vec{m}_1 &=& \vec{f}_1+S_{1,1}\vec{f}_{2}+S_{2,1}\vec{f}_{3}\,,\label{eq:0329cnstr1}\\
	 \vec{m}_2 &=& \vec{f}_1+S_{1,2}\vec{f}_{2}+S_{2,2}\vec{f}_{3}\,, \label{eq:0329cnstr2}
\end{eqnarray}
where $S_{1,1}$ and $S_{2,1}$ are the central values of shape parameters for the 
first Gaussian components, and $S_{1,2}$ and $S_{2,2}$ correspond to the second 
Gaussian components.

\begin{figure}[ht!]
\centering
\includegraphics[width=\columnwidth]{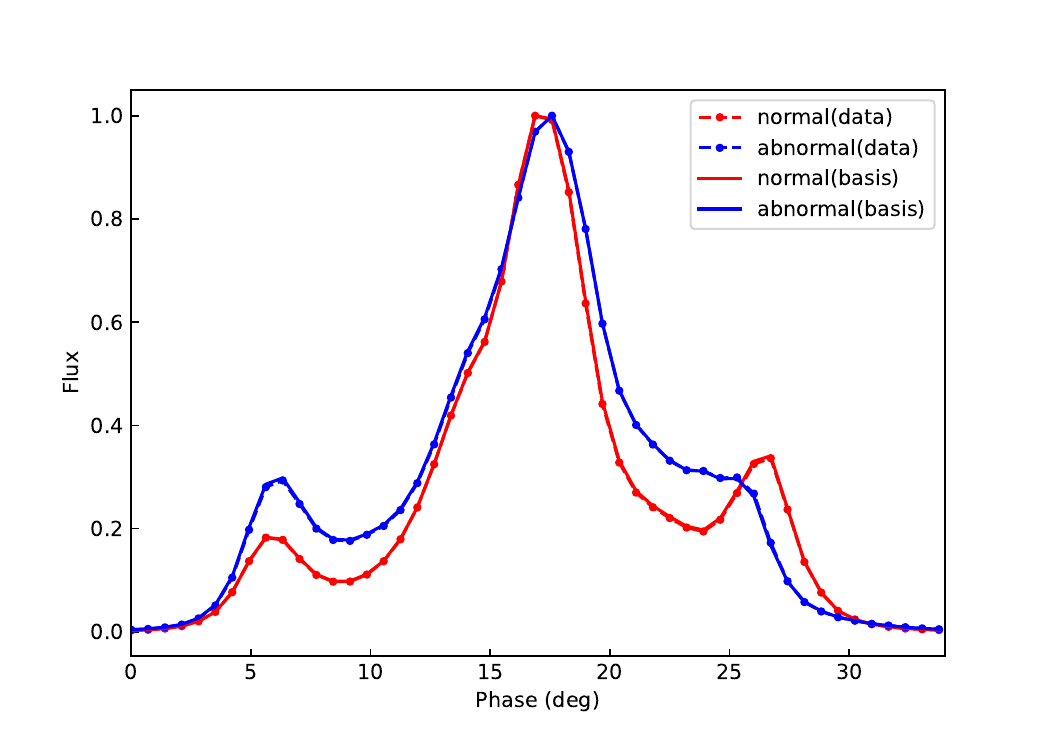}
\caption{Normal and abnormal profiles of \PB~ at 2256 MHz. We 
normalized the profiles by the peak values. The red and blue are for the 
normal and abnormal modes, respectively. The dotted-line curves represent 
the profiles from direct averaging, while the solid curves are the profiles 
produced with our linear decomposition method.
\label{fig:0329prfs}}
\end{figure}

We compare the normal and abnormal profiles obtained by the reconstruction method (\ref{eq:0329cnstr1} and \ref{eq:0329cnstr2}) and by averaging the two modes' pulses in \FIG{fig:0329prfs}.  
Little difference can be noticed. Furthermore, the pulse profiles obtained in our 
analysis are consistent with the results of previous work \citep{cjl2011, 
yz2018}. 

We further measured the percentages of the normal and abnormal modes for 
\PB. We projected all the pulse profiles with S/N$\ge10$ onto the 
profile bases inferred using data with S/N$\ge$50. Then the pulse profiles were
classified again using the GMM method. The measured percentages of the normal and 
abnormal mode is $81.6\pm 0.2$\% and $18.4\pm 0.2$\%, respectively. The results 
are consistent with the previous reports \citep{bartel1982, cjl2011, yz2018}.

%

\begin{figure}[ht!]
\centering
\includegraphics[width=\columnwidth]{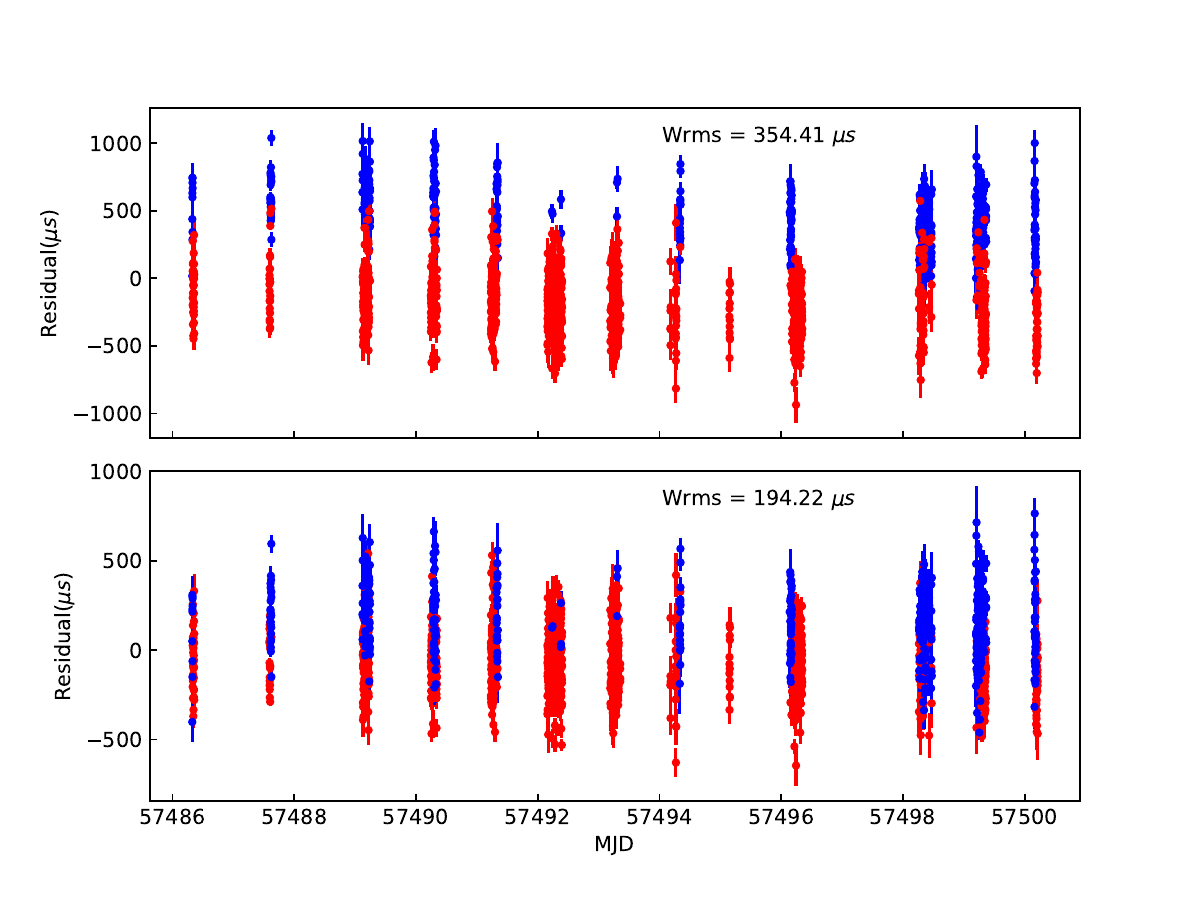}
\caption{Timing results comparison of \PB. Red and blue represent the normal and abnormal modes, respectively. Upper panel: results of timing with TOAs obtained using one overall average pulse as the template. Lower panel: results of timing with TOAs obtained using each mode's average pulse as their own template.}
\label{fig:0329timing}
\end{figure}

Based on the mode classification results, we demonstrated a simple method to improve timing stability for the mode-changing pulsar.
The idea is to reduce the TOA offsets caused by pulse shape variations by using multiple templates in the estimation of TOAs. Focus on studying the mode change induced timing error, we selected two-week long data set with 1,680 one-minute integrations. The short span helps to reduce the influences of red noise or dispersion measure variation on timing residuals.
Then, we integrate the pulse profiles to create the pulse profile templates for timing. The two modes are integrated separately, as the data is classified according to LDM. The two templates are smoothed to avoid the self-noise correlation in TOA determination. We use the two template to compute the pulse TOAs. For comparison, we also obtained the TOAs using the single pulse profile template, which were derived by integrating all data.
The timing results are shown in \FIG{fig:0329timing}.
As one can see, the weighted root mean square (RMS) of the residuals is reduced by approximately 45\%, from 354\,$\mu$s to 194\,$\mu$s.

\subsection{Demonstration with PSR~B0355$+$54}
\label{sec:0355}
\PA~ has a fairly low mode-changing rate. It was observed to exhibit a change in its average pulse profile and linear polarization at a wavelength of 11 cm \citep{MSF80}, while the mode-changing event rate is lower than 5\%. At higher frequencies, the mode change appear to be evident in polarization behaviours. In the polarimetric observation of \PA~ at 10.55 GHz conducted by \citet{XSG95}, the polarization angle exhibited a shift of approximately 45\textdegree, while the shapes of the profiles in normal and abnormal modes barely changed during their observations. Recently, a 100-minute observation at C-band (5.0 GHz) \citep{zhao2019} showed no evidence of mode change.

\begin{figure}[ht!]
\centering
\includegraphics[width=\columnwidth]{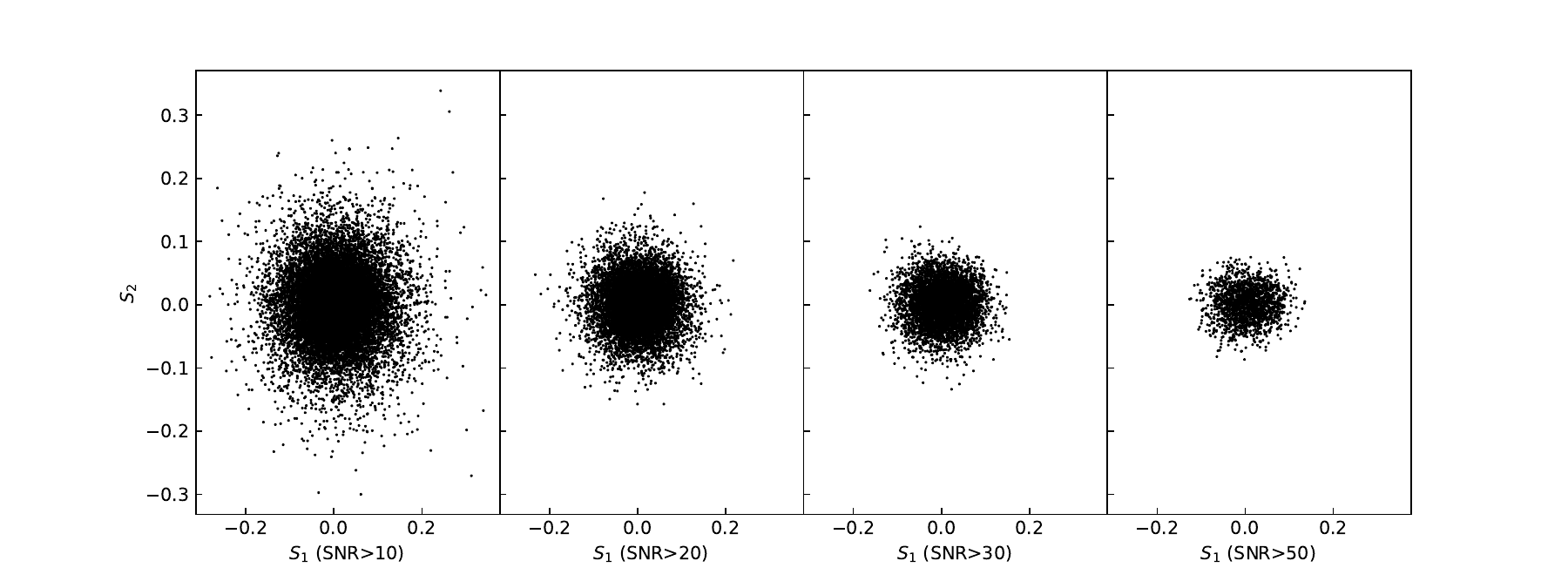}
\caption{Shape parameter values of \PA~ profiles with S/N thresholds 10, 20, 30, 50, 
respectively from the left to the right. The profiles of 4-min integration time is selected 
and the number of the profiles for the four S/N thresholds are 10463, 7123, 4555 and 1892. }
\label{fig:0355A} 
\end{figure}

Following the same procedure as in \SEC{sec:0329}, we obtained the shape parameters for \PA.
The distribution of the shape 
parameters is shown in \FIG{fig:0355A}. There is no distinct 
separation between modes. The shape parameters form a single-component distribution 
unlike the case of \PB~ (i.e., \FIG{fig:0329clss}), where the distribution 
of shape parameters is clearly bimodal. Thus, no clear mode 
changing behavior is detected for \PA.

To check if the modes are diluted by low-S/N profiles, we also plotted the 
shape parameter distribution for samples selected with S/N thresholds of 10, 20, 
30, and 50 in \FIG{fig:0355A}. In none of the cases, we can detect a bimodal 
distribution. Additionally, as we move from the left to the right panels in 
\FIG{fig:0355A}, the S/N increases, 
and the distribution of shape parameters becomes more confined. 
This indicates that the distribution is neither intrinsically bimodal nor affected by the limited S/N.

\begin{figure}[ht!]
\centering
\includegraphics[width=\columnwidth]{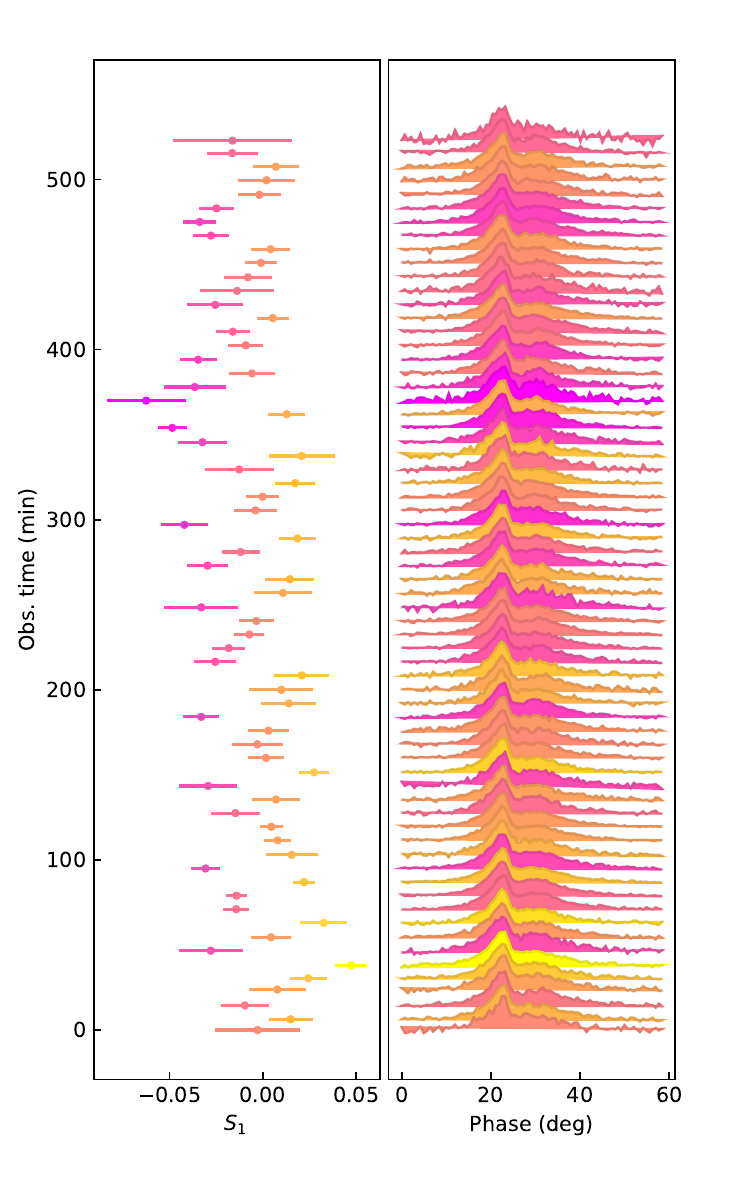}
\caption{The shape evolution of \PA on MJD 57271 on 8 minute interval. Left: 
shape parameters and their errorbars as functions of time. Right: the 
corresponding integrated pulse profiles (8 minute integration).  Colors from 
yellow to purple indicates the value of the shape parameter 
$S_1$.
\label{fig:0355F}}
\end{figure}

\begin{figure}[ht!]
\centering
\includegraphics[width=\columnwidth]{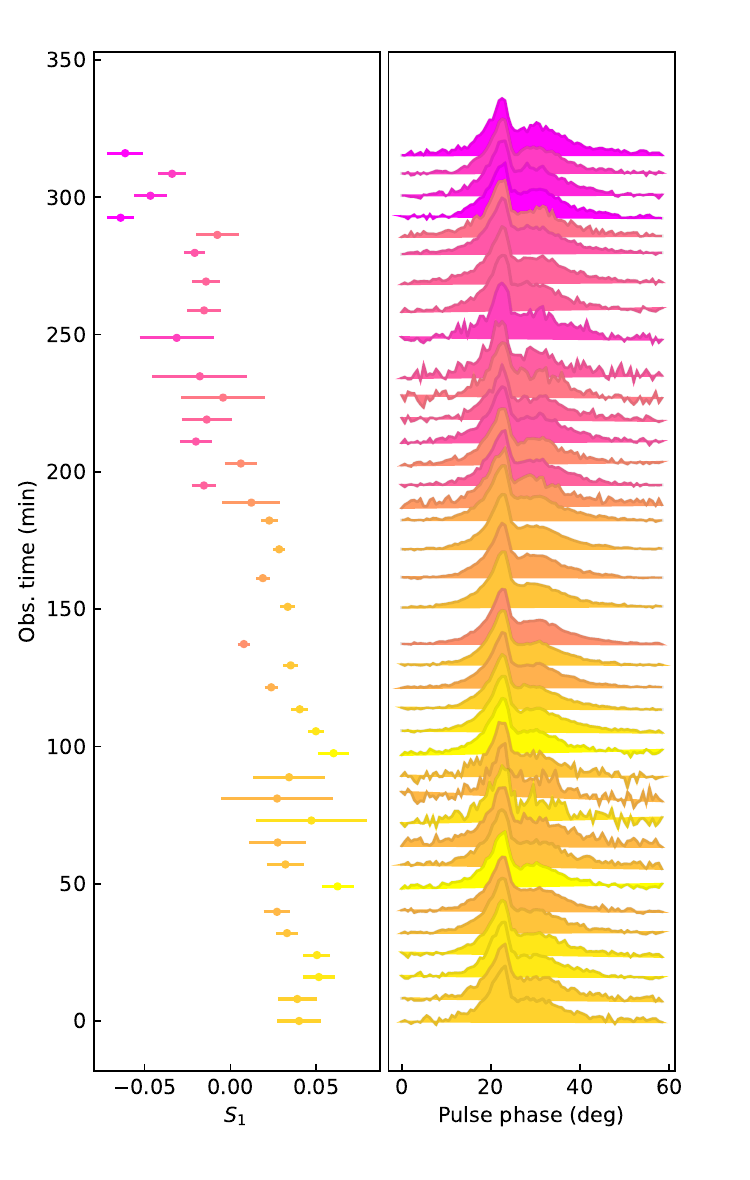}
\caption{The same as \FIG{fig:0355F}, except that the observation is carried out at MJD 59790.
\label{fig:0355G}}
\end{figure}

Despite the absence of a second independent pulse profile mode for \PA, 
the changes in $S_1$ indicates the pulse profile variation. 
For most of the observation 
sessions, the profile variations are rather random in time with an example
given in \FIG{fig:0355F} (MJD 57271). However, for a few 
epochs, we observe systematic evolution of pulse profiles. One such instance (data of MJD 
59790) is illustrated in \FIG{fig:0355G}. It can be seen that the shape parameter $S_1$ 
evolves systematically from positive to negative, accompanied by some fluctuations. The corresponding pulse profiles exhibit more pronounced trailing components.  Clearly, the LDM found the pulse profile variation, and shape parameters is capable of quantifying the phenomenon.

We can compare the current method with more traditional method, the template matching method.
This method is based on the likelihood ratio detector, which maximizes    the 
detection probability for a prescribed false alarm probability \citep{Fisz63}.
As shown in Appendix~\ref{app::chi2}, if the template is known, \emph{the most powerful} statistics to detect 
abnormal mode given the profile templates is to use the statistics $\Delta\chi^2$ 
defined as
\begin{eqnarray}
    \Delta\chi^2\equiv \chi_{\rm n}^2-\chi_{\rm a}^2\,,    \label{eq:dchi2}
\end{eqnarray}
where $\chi_{\rm a}^2$ and $\chi_{\rm n}^2$, as defined in
Appendix~\ref{app::chi2}, represent the $\chi^2$ values of profile residuals obtained by
subtracting the abnormal and normal profile templates from the observed profiles.
A larger value of $\Delta\chi^2$ suggests a better fit of the abnormal mode to the data.

\begin{figure}[ht!]
\centering
\includegraphics[width=\columnwidth]{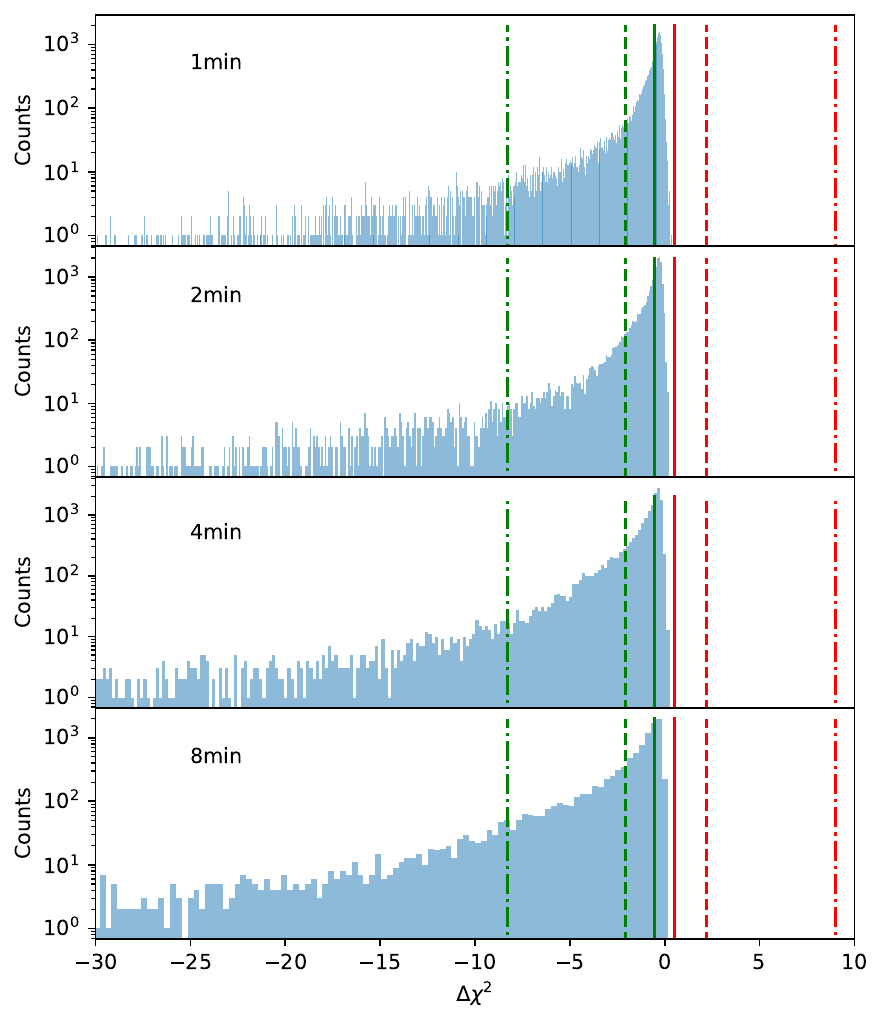}
\caption{Measured $\Delta\chi^2$ distribution for profiles with
S/N$\ge10$ for integration times of 1, 2, 4, and 8 minutes. The
red solid, dashed, and dash-dotted vertical lines would be the
expected mean value of $\Delta\chi^2$, if the observed profile was
in abnormal mode with S/N of 10, 20, and 40. The green solid, dashed,
and dash-dotted vertical lines are the mean value of $\Delta\chi^2$,
if the observed profile is in normal mode. No $\Delta \chi^2$	is high enough 
to reach the expected mean of abnormal mode.  \label{fig:0355dchi2} }
\end{figure}

We utilized the normal and abnormal mode pulse profiles, measured by
\cite{MSF80}, as our templates. 
We restricted our search to pulse profiles with ${\rm S/N} \ge 10$.  The 
distribution of $\Delta \chi^2$ from our
data is shown in Figure~\ref{fig:0355dchi2}. The distribution exhibits
a peak at the expected mean value of $\Delta \chi^2$ for normal mode
profiles at S/N=10. Furthermore, there are no pulses with $\Delta
\chi^2$ values high enough to reach the expected mean for abnormal
mode profiles. Consequently, no abnormal modes were detected, as the
distribution aligns with our expectations for pure normal mode pulse
profiles. Regarding the abnormal mode identified by \cite{MSF80},
a conservative estimation using the 4-minute integrated pulse profile
sample sets a stringent upper limit on the probability of its occurrence,
which is less than $10^{-4}$, given the total number of pulse profile
with ${\rm S/N}\ge 10$ is 10,463.

\section{Discussions and Conclusions}

In this paper, we developed the LDM to study pulsar's pulse shape variations. 
The method is based on the maximum likelihood waveform estimator for waveform, and the iterative method for numerical solution is also invented. 

Mathematically, this LDM is an approach to obtaining the weighted low-rank approximation (WLRA) that commonly obtained through weighted SVD. When no further constraint is implemented, LDM becomes the PCA without data centralization. In LDM, if we set the input parameters to obtain one basis vector per iteration, the solution's uniqueness, orthonormality, and significance-order properties can be automatically guaranteed.
Moreover, there are two primary unique properties of the LDM armed with the iterative algorithm.
First, we can stop the iterative process at a lower rank to save computational time and memory resources. Unlike general SVD or PCA approaches, which always compute the full-rank solution. For pulsar data, only a few bases are generally sufficient for shape analysis.
Second, in the LDM, measurement errors are incorporated into the iterations to weight the data profiles, and formal fitting errors are naturally derived, which is not the case with conventional SVD or PCA methods.

Compared to other methods for analyzing pulsar mode changes, the shape parameters $S_k$  presented in this paper can fully describe the shape of a pulse profile. Given the stability of a pulsar's integrated pulse, the first few orders of the shape parameters contain most of the shape information. This property makes the shape parameters $S_k$ both precise and comprehensive indicators of pulse shape.

To demonstrate the application of the LDM, we applied it to a simulated dataset and real data. It is applied to simulation consisting of 2000 profiles belonging to three distinct shape modes. Based on the method, we studied the pulse variability of \PB~ and \PA~ using KM40 data at 2256 MHz. 
Our results for \PB's mode change events are consistent with previous studies\citep{bartel1982, cjl2011, yz2018}, while we obtain a tighter upper limit for the probability of abnormal mode in \PA.
The rare mode changing phenomenon of \PA~ was reported by \cite{MSF80}. They observed a gradual rather than abrupt shape change in this pulsar, with an occurrence rate of less than 5\% in the available Effelsberg data.

For \PA, we also compared the results with the results of likelihood ratio tests, which confirmed our null detection. Despite that we detect no abnormal mode, we noticed the pulse profile variation from the shape parameter of LDM. We show that the pulse structure of the 
trailing component of \PA~ is changing. For a few epochs, the changing forms a continuous trend.

\begin{acknowledgments}
This work was funded by the National SKA Program of China (Grant No. 2020SKA0120100), the Special Project of Foreign Science and Technology Cooperation, Yunnan Provincial Science and Technology Department (Grant No. 202003AD150010), the National Key R\&D Program of China (Grant No. 2022YFC2205203), the National Natural Science Foundation of China (NSFC, Grant Nos. 12073076, 12173087, 12041303, and 12063003), the CAS `Western Light Youth Project', the CAS-MPG LEGACY Project, and the Max-Planck Partner Group.
\end{acknowledgments}

\appendix

\section{\texorpdfstring{$\chi^2$-}{Chi-square} test and mode-changing detection}
\label{app::chi2}
We define two pulse profile templates for normal and abnormal modes as $P_i$ 
and $P'_i$, respectively. The index $i$ represents the pulse phase, ranging from 1 
to $N$, where $N$ is the total number of data points in the pulse profile. Given 
the observed pulse profile $p_i$ and its corresponding error bar $\sigma_i$, we 
can derive the logarithmic likelihood functions for the observed pulse profile 
under the two hypotheses, that $H_0$: the pulse profile is normal and $H_1$: the 
pulse profile is abnormal. 
\begin{eqnarray}
\log \Lambda &\propto& -\frac{1}{2}\sum_{i=1}^N \left(\frac{p_i -\alpha P_i -\beta}{\sigma_i}\right)^2\,, \\
\log \Lambda' &\propto& -\frac{1}{2}\sum_{i=1}^N \left(\frac{p_i -\alpha' P'_i-\beta'}{\sigma_i}\right)^2.
\end{eqnarray}
Here, $\Lambda$ and $\Lambda'$ represent the likelihoods, i.e. probability 
density functions, assuming that the pulsar is in the normal and abnormal 
states, respectively. In the same notation, $\alpha$ and $\alpha'$ are the 
amplitudes of the observed profile relative to the normal and abnormal 
templates, respectively; and similarly, $\beta$ and $\beta'$ represent the 
baseline levels (DC offsets) for the two modes.

The likelihood ratio test is the most powerful statistic for determining the 
state of the pulsar, i.e., differentiating between the two waveforms with known 
error bars \citep{Fisz63}. It is defined as
\begin{eqnarray}
\Delta \log \Lambda \equiv \log \Lambda - \log \Lambda'\,,
\end{eqnarray}
which is proportional to the $\chi^2$ difference.  Thus, the $\chi^2$ difference 
is also the optimal statistics to detect the mode changing. It is defined as
\begin{eqnarray}
\Delta \chi^2 \equiv \chi^2_{\rm n} - \chi^2_{\rm a},
\end{eqnarray}
where $\chi^2_{\rm a}$ and $\chi^2_{\rm n}$ are
\begin{eqnarray}
\chi^2_{\rm a} = \frac{1}{N}\sum_{i=1}^N\left(\frac{p_i -\alpha' P'_i-\beta'}{\sigma_i}\right)^2\,, \\
\chi^2_{\rm n} = \frac{1}{N}\sum_{i=1}^N\left(\frac{p_i -\alpha P_i-\beta}{\sigma_i}\right)^2\,.
\end{eqnarray}

With the statistics $\chi^2$, in practice, one can compute the $\Delta
\chi^2$ for observed pulse profiles, and determine the state of pulsar
by comparing the measured $\Delta \chi^2$ value with the simulations
under the two statistical hypotheses $H_0$ and $H_1$. It is worth of mentioning that the Wilks's theorem \citep{wilks1938large} does not directly apply here, as the degree of freedom for $H_0$  and $H_1$ are the same here. For our case, one can show that the mean and standard deviation of $\Delta \chi^2 $ are
\begin{eqnarray}
    \langle \Delta \chi^2 \rangle&=&\frac{1}{N}\sum_{i=1}^{N}\left(\frac{\Delta _i}{\sigma_i}\right)^2\,,\\
\sqrt{\langle \Delta \chi^2  \Delta\chi^2\rangle-\langle \Delta \chi^2 \rangle^2}&=&\frac{2}{N}\sqrt{\sum_{i=1}^{N}\left(\frac{\Delta _i}{\sigma_i}\right)^2}\,.
\end{eqnarray}
Here $\Delta_i$ are the difference between the optimal fits using the two profile template, i.e.
$\Delta _i=\alpha P_i +\beta-\alpha' P'_i -\beta'$. In this way, the statistics of $\Delta \chi^2$ also depends on the distribution of S/N. In this paper, we use simulation to determine the average value and threshold of the $\Delta \chi^2$ test.

\bibliography{ms}
\bibliographystyle{aasjournal}

\end{document}